\documentclass[aps,prd,onecolumn,groupedaddress,showpacs,nofootinbib,amssymb,notitlepage]{revtex4-1}
\usepackage{graphics}
\usepackage[pdftex]{graphicx}
\usepackage{tikz}
\usetikzlibrary{decorations.markings}
\usetikzlibrary{decorations.pathmorphing}
\usepackage{amsmath}
\usepackage{amssymb}
\def\be{\begin{equation}}
\def\ee{\end{equation}}
\def\ba{\arraycolsep .1em \begin{eqnarray}}
\def\ea{\end{eqnarray}}

\def\lp{{l_{\rm pl} }}

\def\fig#1{fig.~(\ref{#1})}

\def\s0#1#2{\mbox{\small{$ \frac{#1}{#2} $}}}
\def\0#1#2{\frac{#1}{#2}}

\newcommand{\bit}{\begin{itemize}}
\newcommand{\eit}{\end{itemize}}

\begin{document}

\title{Cosmic Censorship in Quantum Einstein Gravity}
\author{A. Bonanno}
\affiliation{INAF, Osservatorio Astrofisico di Catania, via S. Sofia 78, I-95123 Catania, Italy}
\affiliation{INFN,  Sezione di Catania,  via S. Sofia 64, I-95123, Catania, Italy.}
\author{B. Koch}
\affiliation{Instituto de F\'isica, Pontificia Universidad Cat\'olica de Chile,
Av. Vicuna Mackenna 4860, Santiago, Chile}
\author{A. Platania}
\affiliation{INAF, Osservatorio Astrofisico di Catania, via S. Sofia 78, I-95123 Catania, Italy}
\affiliation{INFN,  Sezione di Catania,  via S. Sofia 64, I-95123, Catania, Italy.}
\affiliation{Universit\`a di Catania, via S. Sofia 63, I-95123 Catania, Italy}

\begin{abstract}
We study the quantum gravity modification of the Kuroda-Papapetrou model induced by the running of the 
Newton's constant at high energy in Quantum Einstein Gravity. 
We argue that although the antiscreening character of the gravitational interaction favours the 
formation of a naked singularity, quantum gravity effects turn  the classical singularity 
into a ``whimper" singularity which remains naked for a finite amount of advanced time. 
\end{abstract}


\maketitle

\section{Introduction}
Spacetime singularities signal the breakdown of the classical description
within General Relativity. Such singularities typically appear at the beginning 
of the universe
and at the endpoint of a gravitational collapse.
In the latter case, after the star's irregularities have been radiated away and the external
field relaxes to its Kerr-Newman geometry, the structure of the interior solution
is rather uncertain, even classically. According to the still unproven cosmic censorship
hypothesis the singularity at $r=0$ in a physical black hole ought to be spacelike and 
its ``evolution" is  described by a generic  mixmaster solution. 
In particular it is believed that  when the curvature reaches Planckian values
quantum fluctuations of the geometry have a self-regulator effect and can eventually  halt the collapse.

An interesting possibility, already proposed  in the context of minisuperspace models \citep{hh83,vilenkin88} coupled to matter fields,
is a transition to a de Sitter core of size $\hbar G \equiv \lp$ \citep{frolov,irina},
a  scenario recently emerged also in the framework of  Asymptotically Safe (AS) gravity  \citep{br00,ko14} 
and loop quantum gravity \cite{rovi14,delo15}. 
On the other hand,  the dynamical role of quantum fluctuations of the metric 
in the formation of the event horizon (EH) and of the central singularity is still unknown.

In contrast to the cosmological singularity, a black hole singularity at $r=0$ develops in time 
because the radial coordinate is time-like inside of the black hole (BH) nucleus. The classical evolution 
is essentially determined by two effects: the radiation from the collapsing star, and the backscattered
inflow of gravitational waves which takes the form $v^{-4l-4}$, where the advanced time $v\rightarrow\infty$
and $l$ is the spherical harmonic multipole of the perturbation \cite{gundlach94}. 
The latter is dominant in the interior only at large advanced times because any perturbation of the metric $\delta g$
will decay as $v^{-2 l-2}$ near $r=0$ \citep{poisson88}. 
The details of the collapse are therefore essential for the understanding
the structure of the singularity at early advanced time, at least for a spherical BH with small asphericities.

In this work we study possible quantum gravity modifications occurring at the early part of the singularity,
by linking the dynamics of the collapse with the structure of the singularity in an quantum-corrected BH interior.
An attempt to answer to the above issue can be pursued within the framework of AS gravity \cite{lare,nire,cope}, 
in particular in the case of  Quantum Einstein Gravity (QEG) \citep{re98,bore,resa}. 

{The basic input of this approach is the existence of a Non-Gaussian Fixed Point (NGFP) 
of the Renormalization Group (RG) flow of the gravitational interaction that allows the definition of  a new
set of relevant operators with a well behaved ultraviolet (UV) limit. }
In QEG the antiscreening character of the gravitational constant
at high energies  has a physically clear interpretation in terms 
of ``paramagnetic dominance" of the vacuum~\citep{nink13},
and it is possible to encode the running of $G$ in a self-consistent manner, 
by renormalization group improving the Einstein equations, 
as explained in cosmological contexts  \cite{br02,fra05,boes,br07,dodo15}. 

Recent studies have modeled the collapse in terms of 
an homogeneous interior surrounded by a Renormalization Group (RG) improved Schwarzschild exterior \citep{torres14,torres14b}.
In this work we instead focus on the dynamical process of BH formation and discuss the structure of the singularity in 
a quantum gravity corrected Kuroda-Papapetrou (KP) model \cite{kuroda,papa}.
We shall show that a violation of the  Cosmic Censorship conjecture is in general {\it favoured} by quantum
gravity, although
the final singularity at $r=0$ turns out to be a rather weak, ``whimper" singularity, 
according to the the Tipler classification \cite{tipler77}.  We argue that, at least in spherical symmetry our result should be 
a generic outcome of the gravitational collapse in QEG.

The rest of this paper is organized as follows. In section II we introduce the basic ideas and we find a RG improved metric describing the gravitational collapse within the VKP model. The general solution of the geodesic equation is studied in the section III, while in section IV we focus on the analysis of the ``nature'' of the singularity in 
comparison to the classical case. Finally, in section V we summarize our main results and comment on possible extensions and applications.

\section{Basic Equations}
In the framework of the RG flow for gravity,  the  ``emerging"  geometry  at the energy scale $k$
can be obtained by solving the RG-improved Einstein's equations
\be\label{effective}
R_{\mu\nu}[\langle g \rangle_k] - \frac{1}{2} R\,\langle g_{\mu\nu}\rangle_k = \Lambda(k) \, \langle g_{\mu\nu}\rangle_k + 8\pi G(k) \, \langle T_{\mu\nu}\rangle_k
\ee
which can be derived from the effective average action for gravity 
$\Gamma_k[g_{\mu\nu}]$ 
via 
\be
\frac{\delta \Gamma_k}{\delta g_{\mu\nu}(x)}[\langle g\rangle_k]=0 . 
\ee
Here, $\Gamma_k[g_{\mu\nu}]$  is a free energy functional which depends on the metric and a momentum scale $k$
which is usually considered as an  infrared cutoff. This functional is similar to the ordinary effective
action which is reproduced in the $k\rightarrow 0$ limit. As the functional $\Gamma_k$ gives rise to an effective
field theory valid near the scale $k$, when evaluated at tree level it describes all quantum gravitational
phenomena at tipical momentum $k$.
In the case of QEG, the functions $G(k)$ and $\Lambda(k)$ are  computed from the exact 
flow equation for gravity \citep{martin} from the ansatz, 
\be
\Gamma_k[g]=(16\pi G(k))^{-1} \int \sqrt{g}\, \{ -R(g) +2\Lambda(k)\}\,d^4 x.
\ee
Clearly, the dynamical implications of a running Newton's constant 
can only be fully exploited once a prescription $k\rightarrow k(x^\mu)$ is provided  to close the system  \eqref{effective}.

In the case of pure gravity it has been argued  that the correct cutoff identification is 
provided by the proper distance of a free-falling observer radially falling into a BH (at least in spherical geometry \cite{br00}).
On the other hand if matter is present in the system the ultraviolet structure of the renormalized flow can be
strongly deformed at high temperatures. 
It is known from scalar field theory \cite{liaos95,liaos98} that
the renormalization scale $k$ and the temperature $T$ play, up to logarithmic corrections, a similar role in determining the scaling laws 
so that $k\sim T$.

In the case at hand, as the BH interior is  homogeneous, it is natural to assume that near the singularity the matter can be 
described by an equation of state consisting of a fluid with radiation type energy density $\rho$, so that  
\be
\label{ci}
\rho \sim T^4 \sim k^4.
\ee
This relation can then be inverted and, for actual calculation, 
we can write $k = \xi \sqrt[4]{\rho}$ where $\xi$ is a positive constant whose precise value
does not qualitatively change the conclusions of our discussion, as we shall see. 
{It is important to remark that this is the only possible choice compatible with a conformally invariant theory
at the NGFP, as gravity is supposed to be.  The introduction of a generic functional form  of the type $k=k(\rho)$ 
would imply the presence of other mass scales not allowed at the NGFP.}

In order to solve \eqref{effective}  it is necessary to resort to some approximation. As discussed in \cite{br00}  a first step 
is to construct a  self-consistent solution which interpolates from the classical one 
expected at infra-red scales $k\rightarrow 0$ and the full quantum corrected geometry expected at 
ultra-violet scales $k\rightarrow\infty$. 
Our strategy in this work therefore amounts to use an iterative procedure: 
we shall first  RG-improve the classical Vaidya solution 
with the cutoff identification \eqref{ci} and then
compute the new quantum corrected stress-energy tensor  following \eqref{effective}. 

Let us write the (average) metric of a spherically symmetric collapsing object as a generalized Vaidya metric \cite{genvaidya} of the type
\begin{equation}\label{vvv}
\langle ds^2 \rangle_k=-f_k(r,v)\,dv^2+2\,dv\,dr+r^2\,d\Omega^2,
\end{equation}
where $f_k(r,v)=1-2 M_k(r,v)/r$ is the RG-improved lapse function.  
The non-zero components of the Einstein tensor
read
\be
G_{vr} =  -\frac{2\,}{r^2} \frac{\partial M_{k}(r,v)}{\partial r},  \;\;\;\;\;\;\;\;
G_{vv}= {\frac{2}{r^2}{\frac {\partial M_k \left( r,v \right)}{\partial v}} +\frac{2\,f_k(r,v)}{r^2}{\frac {\partial M_k \left( r,v \right)}{\partial r}}} 
\ee
\be
G_{\theta\theta}=r\,\frac{\partial^2 M_k(r,v)}{\partial r^2},\;\;\;\;\;\; 
G_{\phi\phi}=r\,{\frac {\partial ^{2}M_k \left( r,v\right)}{\partial {r}^{2}}}\,(\sin\theta)^2.
\ee
As discussed in \cite{goswami,joshi} the energy-momentum tensor has a null-part and non-null part so that 
\be
T_{\mu\nu}=\rho\,l_\mu l_\nu +(\sigma+p)(l_\mu n_\nu + l_\nu n_\nu)+ p g_{\mu\nu},
\ee
where $n_\mu l^\mu = -1$, $l_\mu l^\nu=0$ and 
\be
\rho (r,v) = \frac{1}{4\pi G_0 r^2} \, \frac{\partial M_k(r,v)}{\partial v}
\ee
and
\be
\sigma(r,v) = \frac{1}{4\pi G_0 r^2} \, \frac{\partial M_k(r,v)}{\partial r} \;  , 
\qquad\quad p(r,v) = - \frac{1}{8\pi G_0 r} \, \frac{\partial^2 M_k(r,v)}{\partial r^2}.
\ee

{Classically  $M_{0}(r,v)={G_0 \, m(v)}$,
being $G_0$ the usual Newton's constant and $m(v)$ a dynamical mass function which is a function of the advanced time $v$. 
In this case, as first discussed in \cite{vaidya51,vaidya66}, the stress energy tensor is
that of a null fluid and the classical field equations simply read  $\dot{m}(v)/4\pi r^2 = \rho(r,v)$. 
In the iterative RG-improvement approach the
energy density is  computed at the bare (classical) level and the quantum corrected lapse function is then obtained via the substitution
\begin{equation}\label{runG}
G_0 \;\; \rightarrow \;\; G(k)\,\equiv\,\frac{G_0}{1+\omega \, G_0 \, k^2}, \qquad\qquad  M_0(r,v) \;\; \rightarrow \;\; M_k(r,v) = G(k)\,m(v),
\end{equation}
where $\omega=1/g_* \approx 1.5$,  $g_*$ is the value of the dimensionless running Newton's constant at the 
NGFP\footnote{The existence and the positivity  of the NGFP depend 
in general on the matter content of the system \cite{dpematter}. 
In our iterative approach we use the expression of $G(k)$ obtained from RG methods within the Einstein-Hilbert truncation,
and neglect any possible modification coming from the addition of matter fields.},
and  $k$ is the infrared cutoff defining the running function $G(k)$ 
(note that $G(0)\equiv G_0$). The scaling of the Newton's constant as a function of the energy 
scale $k$ given in eq. \eqref{runG} is obtained by solving the beta functions within the Asympototic Safety Scenario for Quantum Gravity \cite{br00}.}

As we discussed before, $k=k(r,v)$ can be written in terms of the energy density of the ingoing radiation via an equation of state for pure radiation. 
Let us assume that the ingoing flow of radiation is injected in the spacetime at $v=0$ and then stopped at $v=\bar{v}$. 
The mass of the collapsing object thus increases until at $v=\bar{v}$ the accretion stops, and it can be written as
\begin{equation}
m(v)=\begin{cases}0 & v<0 \\ \lambda v & 0\leq v < \bar{v} \\ \bar{m} & v\geq \bar{v}\end{cases}. \label{KPmodel}
\end{equation}
The description of the classical gravitational collapse throught the mass function given in \eqref{KPmodel} 
constitutes the Vaidya-Kuroda-Papapetrou (VKP) model \cite{vaidya66,kuroda,papa}. 

In the RG improved model, for $0\le v<\bar{v}$ the lapse function has the following form 
\begin{equation}\label{vaim}
f_k(r,v)=1-\frac{2\lambda{G_0} v}{r+\alpha \sqrt{\lambda}}\;\; , \qquad{\mbox{with}}\;\;\;   \alpha=\frac{\xi^2 G_0}{\sqrt{4\pi}g_*},
\end{equation}
while for $v\geq \bar{v}$ it continuosly matchs with a quantum corrected Schwarzschild spacetime with 
\be
f_k(r)=1-\frac{2\,\bar{m}\,{G_0}}{r+\alpha \, \sqrt{\lambda}}.
\ee
Therefore the RG-improvement gives rise to a generalized Vaidya-Kuroda-Papapetrou (VKP) model whose mass function reads
\begin{equation}\label{quantumass}
M_k(r,v)=\frac{G_0\,r}{r+\alpha\sqrt{\lambda}} \, m(v),
\end{equation}
where the mass $m(v)$ is the one of the classical VKP-model, given eq. \eqref{KPmodel}. As expected, the continuity of the RG-improved mass function $M_k(r,v)$ along the $v=\bar{v}$ light-cone implies that the quantum corrected spacetime doesn't converge to ``pure" Schwarzschild, but only approaches it asymptotically 
(for $r \rightarrow \infty$). 
The energy momentum tensor for $v\geq\bar{v}$ is characterized by an effective energy density and pressure given by 
\be
\sigma(r) = \frac{\lambda\,\bar{v}}{4\pi r^2} \, \frac{\alpha\sqrt{\lambda}}{(r+\alpha\sqrt{\lambda})^2} \;  , 
\qquad\quad p(r) = \frac{\lambda\,\bar{v}}{4\pi r} \, \frac{\alpha\sqrt{\lambda}}{(r+\alpha\sqrt{\lambda})^3}
\ee
respectively.  A similar situation is encountered in the RG-improved Schwarzschild solution of  \cite{br00},
where the RG-improved Schwarzshild solution is also not a solution of vacuum Einstein equation. 

From equation \eqref{vaim} it is clear that if $\alpha\neq 0$ (for $\alpha=\xi=0$, $f_k$ reduces to the classical lapse function) the improved lapse function $f_k(r,v)$ is well defined in the limit $r\to0$ 
\begin{equation}
\lim_{r\to0}\,f_k(r,v)=1-\frac{\sqrt{16\pi\lambda}}{\omega\,\xi^2}\,v
\end{equation}
and thus the metric is regular in $r=0$. We have to remark that although the metric is regular at $r=0$, this hypersurface is actually singular. The Ricci curvature $R$ and the Kretschmann scalar 
$K=R_{\alpha\beta\gamma\delta}R^{\alpha\beta\gamma\delta}$ are divergent in $r=0$, in particular
\be
R=-\frac{G_0\sqrt{\lambda} v }{\alpha r^2}+O(1/r), \;\;\;\; K=\frac{16 G_0\sqrt{\lambda} v}{\alpha^2 r^4}+O(1/r^3).
\ee 
Nevertheless, it should be noticed that the $r\rightarrow 0$ behavior is less singular than in the classical case, for which $K\sim 1/r^6$.  

The zeros of the lapse function $f_k(r,v)$ identify the so called apparent horizons (AH). In the classical case the apparent horizon is described by the equation $r_\text{AH}(v)=2\,m(v)\,G_0$, while for the RG improved Vaidya metric it is shifted by the constant $\alpha\,\sqrt{\lambda}$
\begin{equation}
r_\text{AH}(v)=2\,m(v)\,G_0-\alpha\,\sqrt{\lambda}\,=\,2\,m(v)\,G_0-\frac{G_0\,\xi^2}{g_*}\,\sqrt{\frac{\lambda}{4\pi}},
\end{equation}
with the condition $r_\text{AH}(v)\geq0$. At the end of the process the apparent and  event horizons (discussed in the next section)
must converge to the improved Schwarzchild radius $r_S= 2\,\bar{m}\,G_0-\alpha\sqrt{\lambda}$. The final mass $\bar{m}$ of the BH, 
the value of $\bar{v}$, and the value of $\lambda$ are thus related by the condition $\bar{m}=\lambda\bar{v}$. Furthermore, since the Schwarzchild radius $r_S$ must be positive, there exists a minimum period $\bar{v}$ of irradiation, necessary to form a black hole
\begin{equation}
r_S= 2\,\lambda\bar{v}\,G_0-\alpha\sqrt{\lambda}\geq0 \qquad \Rightarrow \qquad 
\bar{v}\geq v_\text{min}(\lambda)\equiv \frac{\xi^2}{2\,\,g_*}\,\sqrt{\frac{1}{4\pi\lambda}}.
\end{equation}
Here, the minimum value $v_\text{min}$ is a function of the radiation rate $\lambda$.

\section{Outgoing radial null geodesics}
\subsection{Classical VKP model}

{It is useful to remind the basic properties of the classical VKP model \cite{joshiKP1,joshiKP2}}.
In the Vaidya geometry outgoing radial light rays are represented 
as solutions of 
\begin{equation}
\frac{dr}{dv} =\frac{1}{2}\left(1-\frac{2 G_0 m(v)}{r}\right) \label{vaidya} .
\end{equation}
If  $m(v) = \lambda v$ the evolution of the spacetime is halted at $r=0$ 
by a strong spacelike singularity \cite{kuroda}, as it will be discussed in the next section. 
In particular it is possible to show that 
if $\lambda> \lambda_c=\tfrac{1}{16\,G_0}$ the singularity in $r=0$ is always covered by an event horizon. 
On the other hand, if $\lambda\leq\lambda_c$ two linear solutions of eq.(\ref{vaidya}),  $r_\pm(v)=\mu_\pm\, v$ with 
\begin{equation}
\mu_\pm=\frac{1\pm\sqrt{1-16\, \lambda \,G_0}}{4}\; , \label{mupm}
\end{equation}
appear.  The positive one is the Cauchy horizon, the other one is the tangent (in the $(r(v),v)$ space) to the event horizon at the ``point"  $(r=0,v=0)$. In this case the singularity is globally naked: all the solutions belonging to the region between the two linear solutions represent light rays that start 
from the singularity $(r=0,v=0)$ and will reach  the observer at infinity. In particular the Cauchy horizon is the first light ray escaping from the singularity. 
 
\begin{figure}
\includegraphics[scale=0.3]{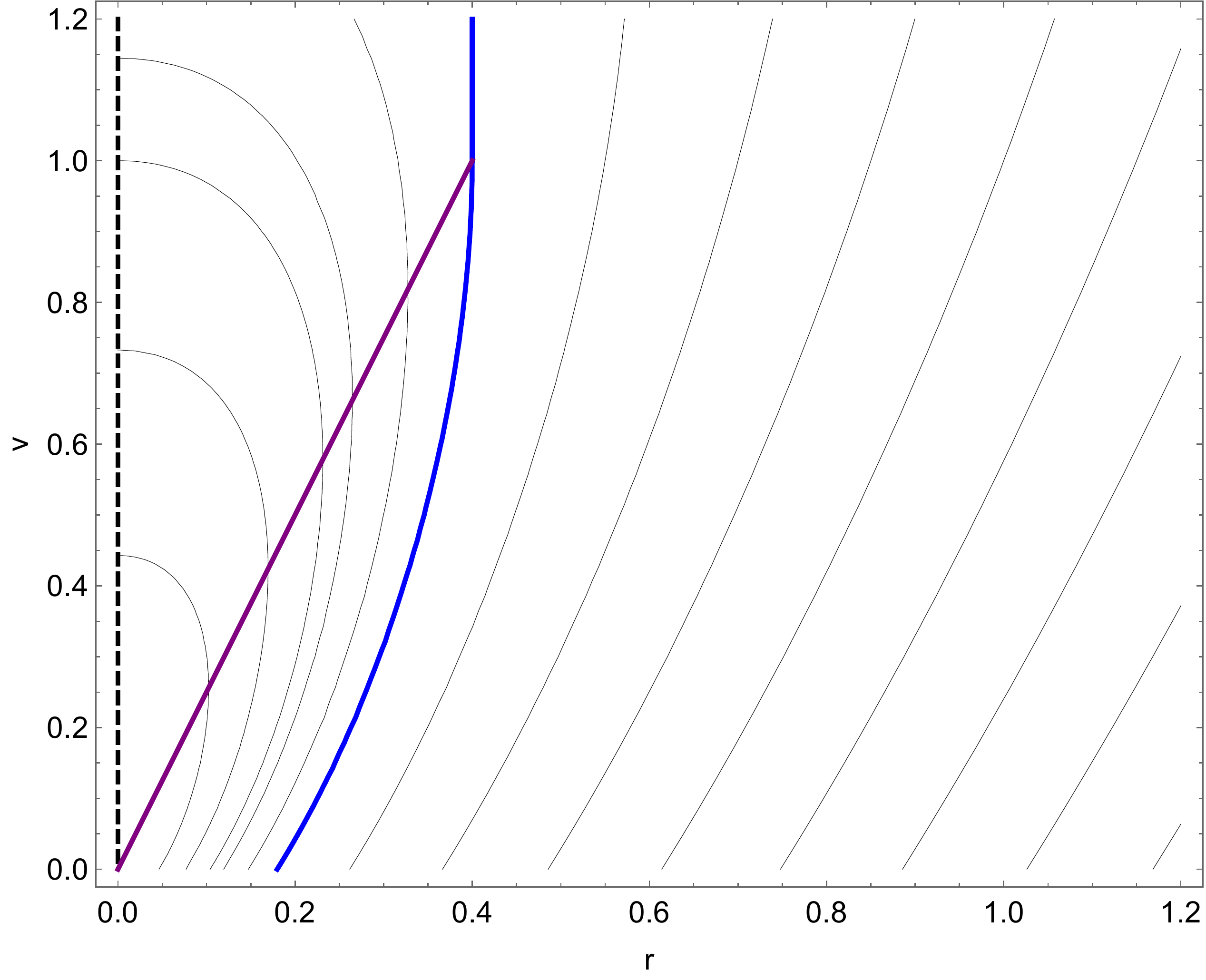}
\includegraphics[scale=0.3]{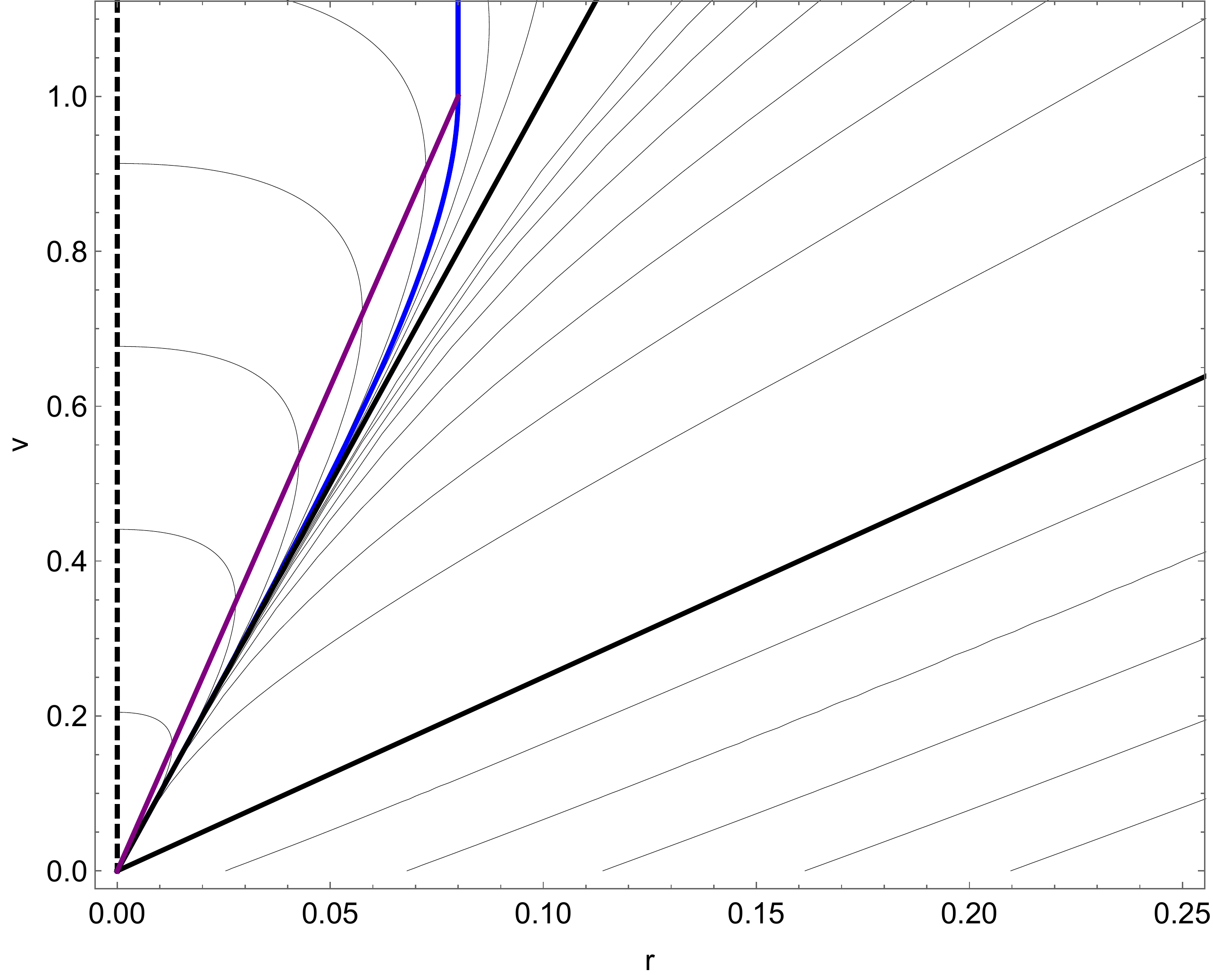}
\caption{Phase diagram $(r(v),v)$ in the classical VKP model. The blue line is the event horizon (EH), the purple line is the apparent horizon (AH), and the black lines represent the family of solutions of the classical geodesic equation. Left panel: For $\lambda>\tfrac{1}{16\,G_0}$ the singularity in $r=0$ is covered by the event horizon. Right panel: In the case $\lambda\leq\tfrac{1}{16\,G_0}$ a Cauchy horizon $r_+(v)=\mu_+ v$ (external bold black line) is formed, and the singularity is naked.}
\label{figclass}
\end{figure}

The general solution to \eqref{vaidya} reads
\begin{equation}
-\frac {2\,\text {ArcTan}\left[\frac {v - 4 r(v)} {v\sqrt {-1 + 16\,\lambda\,G_0}} \right]} {\sqrt {-1 +16\,\lambda\,G_0}}+2\,\text{log}(v)+\text{log}\left[ 2\lambda\,G_0 - \frac{r(v)}{v}+2\,\frac {r^2(v)} {v^2} \right]=C \label{sol1}
\end{equation}
where  $C$ is an arbitrary integration constant. A set of this family of solutions, obtained by varying the constant $C$, is shown in \fig{figclass}. 
One finds that the family of solutions of \eqref{vaidya} for  $\lambda\leq\lambda_c$ can be described by the following implicit equation
\begin{equation}
\frac{|r(v)-\mu_- v|^{\,\mu_-}}{|r(v)-\mu_+ v|^{\,\mu_+}}=\tilde{C}
\end{equation}
where  $\tilde{C}$ in an arbitrary complex constant \citep{werner85}. 
Note that the linear solutions $r_\pm(v)$ are recovered as  particular cases of  this implicit equation. 

\subsection{RG-improved VKP model}
If we now encode the leading quantum correction to this model by using \eqref{vaim} to form the new Vaidya line element, 
the equation for the outgoing radial null geodesic reads
\begin{equation}
\dot{r}(v)=\frac{1}{2}\left(1-\frac{2\,\lambda v\,G_0}{r(v)+\alpha \, \sqrt{\lambda}}\right) \label{eqi},
\end{equation}
where it can be noticed that the effect of a running gravitational coupling is to shift $r(v)$ by a factor $\alpha \, \sqrt{\lambda}$. 
In particular, if $\lambda\leq\tfrac{1}{16\,G_0}$ the solutions of \eqref{eqi} are implicitly  defined by 
\begin{equation}
\frac{|r(v)+\alpha \, \sqrt{\lambda}\,-\mu_- v|^{\,\mu_-}}{|r(v)+\alpha \, \sqrt{\lambda}\,-\mu_+ v|^{\,\mu_+}}=\tilde{C}
\end{equation}
and  the two linear solutions are now given by
\begin{equation}
r_\pm(v)=-\alpha \, \sqrt{\lambda}+\mu_\pm \,v \label{rgcsol},
\end{equation}
with $\mu_\pm$ defined in  \eqref{mupm}. 
These solutions are obtained only for $\lambda\leq\tfrac{1}{16\, G_0}$ 
but, {in contrast to the classical case}, the value $\lambda_c$ at which the singularity become globally naked is now greater than $\tfrac{1}{16\, G_0}$. Furthermore, the  solution $r_+(v)$ in eq. \eqref{rgcsol} has no longer the meaning of Cauchy horizon, since it emerges from $r=0$ well after the formation of the singularity, i.e. for $v>0$.

\begin{figure}
\includegraphics[scale=0.4]{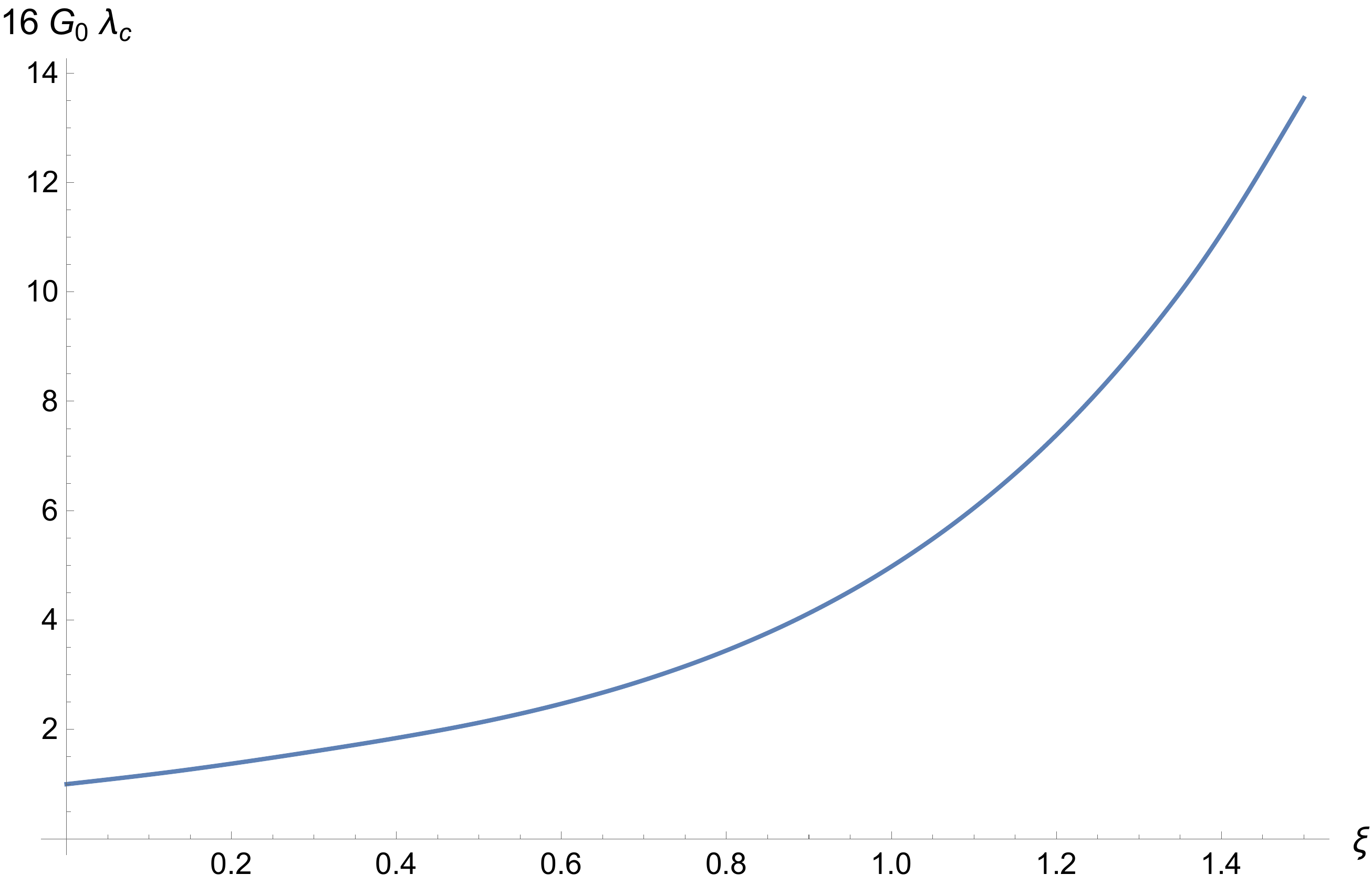}
\caption{Relative critical value $\lambda_c / \frac{1}{16 G_0}$ as a function of $\xi$. As it is clear from the picture, $\lambda_c$ increases monotonically with $\xi$ and in particular it reduces to the classical value when $\xi=0$. }
\label{figxi}
\end{figure}
{In this case  it is not possible to determine analytically the critical value $\lambda_c$ 
but one has to determine it  numerically as the value of $\lambda$
such that the radius of the outgoing null geodesic is zero at $v=0$}. It turns 
out that $\lambda_c$ is always greater than its classical counterpart, and
in particular the dependence of the critical value $\lambda_c$ on the RG parameter $\xi$ is reported in fig. \ref{figxi}.

\begin{figure}
\includegraphics[scale=0.33]{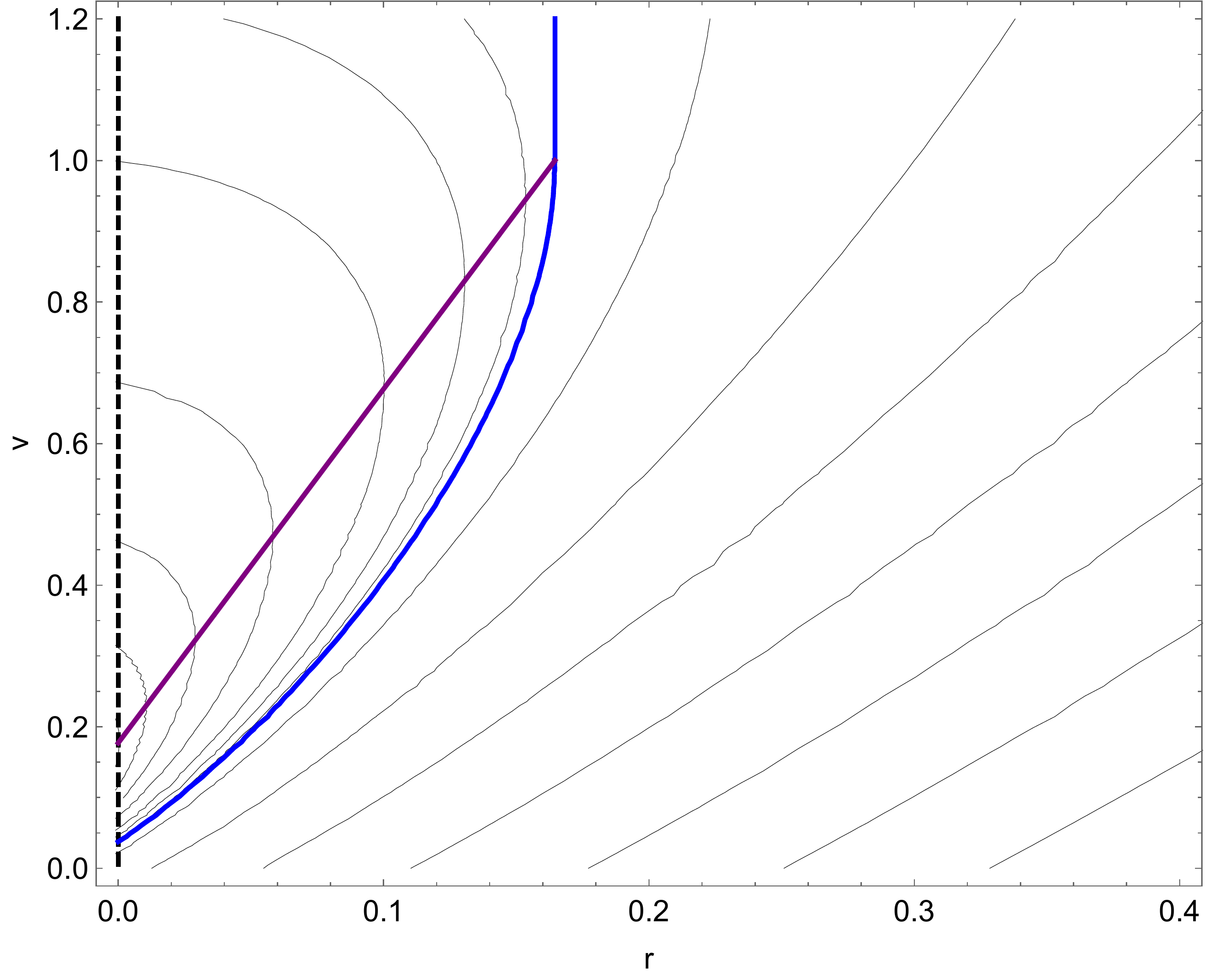}\;\;\;\includegraphics[scale=0.359]{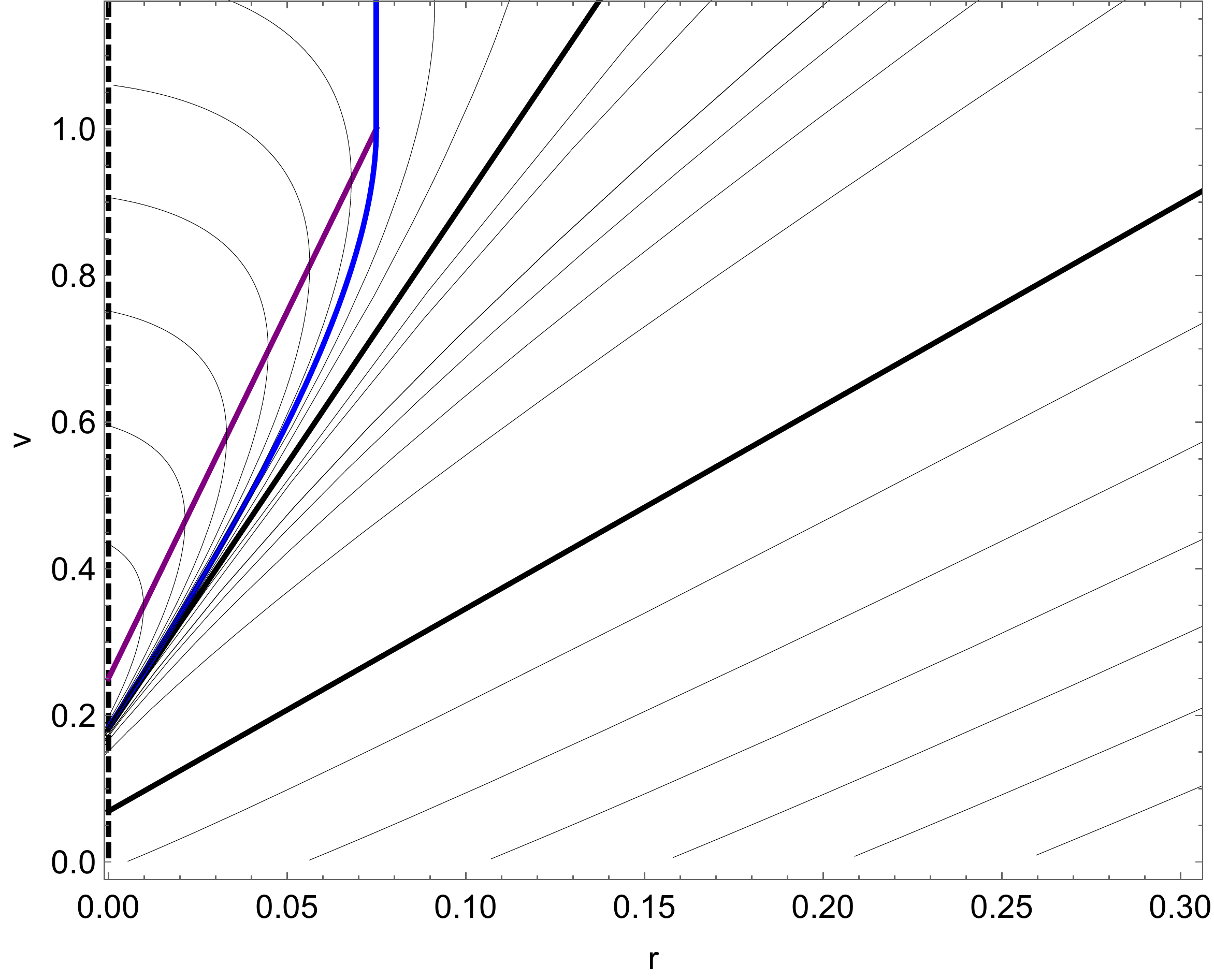}\;\;\;\;\;\;
\caption{Phase diagram $(r(v),v)$ in the improved VKP model, for $\tfrac{1}{16\, G_0}<\lambda\leq\lambda_c$ (left panel) and $\lambda_c\leq\tfrac{1}{16\, G_0}$ (right panel). The blue line is the EH, the purple line is the AH, 
and the black curves are particular solutions of the improved geodesic equation. 
For $\lambda\leq\lambda_c$ the singularity in $r=0$ is globally naked, as the EH forms just after 
the formation of the singularity. {Since in the improved case the affine solution $r_+(v)$ 
has no longer the meaning of a Cauchy horizon, the two cases depicted in these pictures are physically the same}.}
\label{figq}
\end{figure}

By analyzing the solutions of the geodesic equation, we studied the general behavior of the trajectories in 
the phase diagram depending on the value of the radiation rate $\lambda$. 
As it can be seen in the fig. \ref{figq}, for $\lambda\leq\lambda_c$ the singularity is globally naked and moreover 
it extends in time, up to the formation of the event horizon for $v>0$. 
Moreover, as $r{,_\mu} r^{,\mu} = f_k(r,v)$, the singularity $r=0$ is timelike for 
$v<\tilde{v}$, null for $v=\tilde{v}$ and spacelike for $v>\tilde{v}$, 
where $\tilde{v}=\tfrac{\alpha\sqrt{\lambda}}{2\lambda G_0}$ is the value of $v$ at which the improved AH forms, $r_\text{AH}(\tilde{v})=0$. The global structure of the spacetime is described in  fig. \ref{figcon}. 
We can conclude that at least within the VKP model, quantum gravity effects can favor 
the presence of naked singularities. However, as it will be discussed in the next section, the global effect 
of a running Newton's constant is to render the singularity in $r=0$  much milder and integrable. 
\begin{figure}
\centering
\resizebox{0.45\textwidth}{!}{\begin{tikzpicture}
\tikzset{->-/.style={decoration={markings,mark=at position {#1} with {\arrow{>}}}, postaction={decorate}}}
\tikzset{-<-/.style={decoration={markings,mark= at position {#1} with {\arrow{<}}},
postaction={decorate}}}
\draw[decoration={zigzag, amplitude=0.6mm, segment length = 2mm}, decorate,thick] (-3-1-1.5+0.5,4-4+1) -- (-3+3,4-3) node[pos=.5,sloped,above] {$r=0$};
\draw [decoration={zigzag, amplitude=0.6mm, segment length = 2mm}, decorate,thick] (-3-1-1.5+0.5,4-4+1) -- (3.5-1.5-2.5-1.5-1.5,4-4-2.5+1.5+0.5)  node[pos=.52,sloped,below] {$r=0$};
\draw [decoration={zigzag, amplitude=0.6mm, segment length = 2mm}, decorate,thick](3.5-1.5-2.5-1.5-1.5,4-4-2.5+2) -- (3.5-1.5-2.5-1.5-1.5,4-4-2.5-2);
\draw [thick] (3.5-1.5-5.5,4-4-2.5-2) -- (3.5-1.5-5.5,-2.5-4+1.5-1);
\draw (-3+3,4-3) -- (4-2.25,0.5-1.25) node[pos=.35,right] {$\;\mathcal{J}^+$};
\draw (4-2.25,0.5-1.25) -- (3.5-1.5-5.5,-2.5-4+1.5-1) node[pos=.6,sloped,below] {$v \; \longrightarrow$} node[pos=.17,right] {$\,\mathcal{J}^-$};
\draw [dashed] (3.5-1.5-5.5,4-4-2.5-2) -- (3.5-1.5-5.5+0.75,4-4-2.5-2-0.75) node[pos=.5,sloped,above] {$v=0$};
\draw [dashed] (-3.5,-0.5) -- (-3.5+2.75,-0.5-2.75) node[pos=.87,sloped,above] {$v=\tilde{v}$};
\draw [dashed] (-1.5-1.5,-0.5+1.5) -- (0.25,-2.25) node[pos=.885,sloped,above] {$v=\bar{v}$};
\draw [thick] (3.5-1.5-2.5-3,4-4-2.5-3+0.5+0.5) -- (3.5+0.5-2.5-0.5,-2.5-0.5+2.5+0.5) node[pos=.36,sloped,below] {CH};
\draw [thick] (3.5-1.5-2.5-1.5-1.5,4-4-2.5+1.5-1.5) -- (3.5+0.5-2.5-1.5,-2.5-0.5+2.5+1.5) node[pos=.2,sloped,below] {EH};
\draw [thick] (-3.5,-0.5) to [out=270+55,in=270-50] (-1.5,-0.5);
\node at (-2.55,-0.65) {AH};
\end{tikzpicture}}
\caption{Global structure of the spacetime for $\lambda\leq\lambda_c$. \label{figcon}}
\end{figure}

\section{Nature of the singularity in the improved Vaidya space-time}
As it was already noticed, the divergence of Ricci curvature $R$ and the Kretschmann scalar $K$ is different in the case of the Vaidya improved metric.
Let us discuss this issue in detail by following the approach of \cite{goswami}. We write the geodesic equation for null rays of the line element \eqref{vvv} as a dynamical system
\begin{equation}
\begin{cases}\frac{\mathrm{d}v(t)}{\mathrm{d}t}=N(r,v) \\ \frac{\mathrm{d}r(t)}{\mathrm{d}t}=D(r,v)\end{cases} \label{ds},
\end{equation}
where $t$ is a parameter and the functions $N(r,v)$ and $D(r,v)$ are defined as
\begin{equation}
N(r,v)=2\,r \qquad\qquad D(r,v)=r-2\,M_k(r,v).
\end{equation} 
The singularities are the fixed point solutions of the system \eqref{ds} (i.e. $r=0$ and $M_k(0,v)=0$).
The behavior of the trajectories near the singularity can be studied by linearizing the system around the fixed point solution
\begin{equation}\label{dslin}
\begin{cases}\frac{\mathrm{d}v(t)}{\mathrm{d}t}=\dot{N}_\text{FP}\,(v-v_\text{FP})+N'_\text{FP}\,(r-r_\text{FP}) 
\\ \frac{\mathrm{d}r(t)}{\mathrm{d}t}=\dot{D}_\text{FP}\,(v-v_\text{FP})+D'_\text{FP}\,(r-r_\text{FP})\end{cases} .
\end{equation}
Here a ``prime'' denotes differentiation respect to $r$, while a ``dot'' denotes 
differentiation respect to $v$, and the subscript FP means that the derivatives are evaluated at the fixed point (singularity).  
In order to characterize the singularity we have to study the eigenvalues of the stability matrix $J$ of the system (\ref{dslin})
\begin{equation}
\chi_\pm=\frac{1}{2}\left(\text{Tr}J\pm\sqrt{(\text{Tr}J)^2-4\,\text{det}J}\right),
\end{equation}
where
\begin{eqnarray}
&&\text{Tr}J=\dot{N}_\text{FP}+D'_\text{FP}=1-2\,M'_{FP} \\
&&\text{det}J=\dot{N}_\text{FP}D'_\text{FP}-\dot{D}_\text{FP}N'_\text{FP}=4\,\dot{M}_{FP}.
\end{eqnarray}
A singularity is called  \textit{locally naked} if the FP is a repulsive node 
($\text{Tr}J>0$, $\text{det}J>0$, and $(\text{Tr}J)^2-4\,\text{det}J>0$). The strength of the singularity 
is given by
\begin{equation}
S=\frac{\dot{M}_{FP}\,X_{FP}^2}{2},
\end{equation}
where $\displaystyle{X_{FP}\equiv \lim_{(r,v)\to \text{FP}} \tfrac{v(r)}{r}}$ 
is the tangent vector to the radial null geodesic at the FP. The singularity is  \textit{strong} 
if $S>0$, otherwise it is called \textit{integrable}. The tangent vectors $X_{FP}$ defining 
the directions of the trajectories near the singularity are  the non-marginal characteristic directions 
relative to the eigenvalues $\chi_\pm$. One sees that these characteristic directions are given by
\begin{equation}
r(v)=r_{FP}+\frac{\chi_\pm}{2}\,(v-v_{FP}),
\end{equation}
so that $X_{FP}=\tfrac{2}{\chi_\pm}$. In particular if $\text{det}J\neq0$ 
one obtains  the expression $X_{FP}=\tfrac{2\,\chi_\mp}{\text{det}J}$ 
(the same relation found by \cite{goswami} with a different method). 
As an example, we first review the simple Vaidya case, in which the generalized mass
function is $M_0(r,v)=m(v)\,G_0$. 
In this case the singularity is the point $\text{FP}=(0,0)$, and:
\begin{equation}
\text{Tr}J=1-2\,M'_{0}=1 \qquad\qquad \text{det}J=4\,\dot{M}_{0}=4\,\lambda\,G_0 
\end{equation}
thus
\begin{equation}
\chi_\pm=\frac{1}{2}\left(1\pm\sqrt{1-16\,\lambda\,G_0}\right)
\end{equation}
and the point $(0,0)$ is a locally naked singularity if $\lambda\leq\tfrac{1}{16\,G_0}$. 
It should be noticed that the linear solutions $r(v)=\tfrac{\chi_\pm}{2}\,v\equiv \mu_\pm \,v$ 
are  the characteristic directions of the corresponding dynamical system's fixed point. 
In particular in this case $X_{FP}=\tfrac{\mu_\mp}{\lambda\,G_0}$ and therefore $r=0$ is  strong singularity ($\dot{M}_{0}>0$). 

In the improved Vaidya case the generalized mass function is
$M_k(r,v)=m(v)\,G(r)$ with $\lim_{r\to0} G(r)=0$,  
and the fixed points are the solutions of the following system
\begin{equation}
\begin{cases}2\,r=0 \\ r-\frac{2\lambda v \,G_0}{r+\alpha\,\sqrt{\lambda}}\,r=0\end{cases} 
\end{equation}
Therefore a \textit{line of fixed points} is present at $r=0$ and for a given fixed point of the type $(0,v_0)$  
\begin{equation}
\text{Tr}J=1-2\,M'_{FP}=1-\frac{2\lambda v_0\,G_0}{\alpha\,\sqrt{\lambda}},\;\;\;\;\;\;\;\;\; \text{det}J=4\,\dot{M}_{FP}=0.
\end{equation}
In particular, the determinant is zero because the running gravitational coupling $G(r)$ is zero in the limit $r\to0$. 
This fact implies that one of the eingenvalues of $J$ is always $\chi_-=0$, 
while the other one depends on the position of the fixed point in the line $r=0$
\begin{equation}
\chi_+(v_0)\equiv \text{Tr}J=1-\frac{2\lambda v_0\,G_0}{\alpha\,\sqrt{\lambda}}.
\end{equation}
In this case the singularities form a line of improper nodes (in the dynamical system classification).  
The characteristic directions associated with the lines are
\begin{equation}
r=0 \qquad\qquad v=v_0+\frac{2\,r}{\chi_+(v_0)}.
\end{equation}
The only non-marginal direction is the family $X_{FP}=\tfrac{2}{\chi_+(v_0)}$ 
and (assuming that we can compute the strength of the singularity as in the case of a node) 
because of $\text{det}J=4\,\dot{M}_{FP}=0$, in our case the strength of the singularity is zero. 
The improved Vaidya space-time is thus characterized by a {line of integrable singularities} at $r=0$.

{We have to remark that this result does not depend on the cutoff identification. The strength of the singularity is zero as the Newton's constant tends to zero in the ultraviolet limit $k\to\infty$, i.e. for $r\to0$, and thus this is a general result of Asymptotic Safety.}

This result can also be understood in a more intuitive way: for a given dynamical 
system of the type $\dot{\textbf{x}}(t)=\textbf{f}\,[\textbf{x}(t)]$, 
the volume $V(t)$ of a cluster of initial conditions in the phase space $\{\textbf{x}(t)\}$ evolves according to
\begin{equation}
\frac{1}{V(t)}\,\frac{\mathrm{d}V(t)}{\mathrm{d}t}=\nabla\cdot \textbf{f}\,[\textbf{x}(t)]\equiv \text{det}J.
\end{equation}
Therefore, if $\,\nabla\cdot \textbf{f}\neq 0$ $\,\forall t$ then $V(t)\to0$ for $t\to+\infty$ 
(or $t\to-\infty$, depending on the positivity of $\text{Tr}J$), i.e. the trajectories will asymptotically approch a geometrical 
object having zero-measure, as it happens in the classical VKP model. 
In our case $\nabla\cdot \textbf{f}=0$ $\,\forall t$, thus the volume of a cluster of initial conditions 
is always conserved: this fact implies that an object falling into the singularity is not ``destroyed'' 
by the tidal forces (its volume is not reduced to zero) and the singularity  is integrable.

\section{conclusions}

In this paper we have presented a dynamical model of a gravitational collapse based on a RG improved Vaidya 
metric by using the RG equation for QEG.  Our conclusions are that for a generic set of initial 
conditions the singularity is globally naked, even for flow conditions that were not naked in the classical case. 
On the other hand, due to the running of the gravitational coupling
at Planck energies,  the divergence of the Ricci curvature is much milder than in the classical case. 
In particular the singularity is always integrable for any value of the efficiency of the accrection mass scale $\lambda$. 
It must be stressed that  inclusion of a possible running of the cosmological constant in the lapse function would not
change our conclusions as this contribution, with the dynamical cutoff identification proposed in this
paper, is always vanishing in the $r\to 0$ limit
(as long as for $k^{2a}\sim \rho$ with $a>1$). 
It would be interesting to discuss the possible astrophysical consequences of the presented model, 
an issue we plan to discuss in a separate publication.

\section*{Acknowledgements}
A.P and B.K. acknowledge the support of FONDECYT 1120360 and 1161150.

\bibliography{nk}

\end{document}